\documentclass[aps,twocolumn,superscriptaddress,showpacs,pre,
nofootinbib,floatfix]{revtex4}

\usepackage{graphics}
\usepackage{graphicx}
\usepackage{epsfig}

\usepackage{amssymb}

\newcommand{\tresp}{t_{\mbox{\tiny\sf resp}}}

\newcommand{\tbounce}{T_{\mbox{\tiny\sf bounce}}}
\newcommand{\tfirst}{T_{{\mbox{\tiny\sf 1st}}}}
\newcommand{\tmass}{t_{\mbox{\tiny\sf mass}}}
\newcommand{\tfly}{t_{\mbox{\tiny\sf fly}}}
\newcommand{\tgwnue}{t_{\mbox{\tiny\sf GW}}}
\newcommand{\fit}{{\mbox{\sf\tiny Fit}}}
\newcommand{\true}{{\mbox{\sf\tiny True}}}
\begin{document}



\title{Neutrinos from Supernovae as a Trigger for
  Gravitational Wave Search}

\author{G. Pagliaroli}
\affiliation{INFN, Laboratori Nazionali del Gran Sasso, Assergi (AQ), Italy}
\affiliation{University of L'Aquila, Coppito (AQ), Italy}

\author{F. Vissani}
\affiliation{INFN, Laboratori Nazionali del Gran Sasso, Assergi (AQ), Italy}

\author{E. Coccia}
\affiliation{INFN, Laboratori Nazionali del Gran Sasso, Assergi (AQ), Italy}
\affiliation{University of Rome ``Tor Vergata'', Rome, Italy}

\author{W. Fulgione}
\affiliation{Istituto di Fisica dello Spazio Interplanetario (INAF), I-10133 Torino, Italy} \affiliation{INFN, I-10125 Torino, Italy}


\begin{abstract}
Exploiting 
an improved analysis of 
the $\bar{\nu}_e$ signal from the explosion of a galactic core collapse supernova, we show that it is possible to identify within about ten
milliseconds the time of the bounce, which is strongly correlated to the time of the maximum amplitude of the gravitational signal. This allows
to precisely identify the gravitational wave burst timing.
\end{abstract}



\maketitle


\paragraph*{\bf Introduction:} Neutrinos and Gravitational Waves (GW)
are emitted deep inside the Supernova (SN) core and reach terrestrial detectors practically unmodified. They are unique probes to obtain
information on the still puzzling scenario of the explosion, in particular on multi-dimensional dynamics of the proto-neutron star and on the
physics of the postshock region.

GW have not been observed directly yet \cite{Virgo,Kawabe:2008zz,igec},
but detectors of enhanced sensitivity will operate
in the forthcoming years. One of their aims is  
the search of GW bursts from core collapse SNe.
An external trigger can be a very precious (or just necessary) tool for a successful search of such an impulsive sources of GW. In fact, an
external trigger permits GW detectors to lower the event detection threshold, reaching a higher detection probability at a fixed false alarm
probability~\cite{Abbott:2008mr,Antonioli:2004zb,Fleurot:2007zz}.

The neutrino signal from a core collapse supernova has been detected
for SN1987A. 
Despite low statistics and doubts, 
it can be said that these observations
are in overall agreement with the expectations~\cite{bahcall,astr}; even more, they
provide some support~\cite{ll,noi} to the existence of a brief phase of intense neutrino luminosity expected in the standard scenario. Several
detectors are ready to detect the future galactic SN and to test into
details the physics of the explosion; when this signal will be available,
using a proper analysis procedure, it will presumably become {\em the} external trigger for the search of a GW burst.

In this work, we quantify
the potential of this type of trigger, 
making reference to existing neutrino and GW detectors. We show that it is possible to predict precisely the time window for GW search by
analyzing the neutrino signal from a future galactic supernova. We argue that the size of this window, dictated by astrophysics, can be matched
to the duration of the GW signal itself, that is several orders of magnitude smaller than the duration of the neutrino emission.




\paragraph*{\bf Time relation between GW and neutrinos:}
GW can be emitted during the collapse, or during the explosion, of a core collapse SN due to the star's changing mass quadrupole moment. Recent
simulations \cite{Ott2007,Dimmelmeier2007,Dimmelmeier:2008iq} show that this gravitational signal is emitted when the homologous collapse of the
inner core halts, as dictated by the stiffening of the equation of state at nuclear density,
and the bounce is \emph{pressure-dominated} without strong influence of the rotation. 
Therefore, it is possible to define a generic GW waveform which exhibits a positive pre-bounce rise and a large negative peak, followed by a
ring-down; the time of the bounce is strongly correlated to the time of the maximum amplitude of the gravitational signal \cite{FryerNew2003}.

The duration of this signal is about $10$ ms. Therefore, our goal is the identification of the time of the bounce with an error of the same
order using the neutrino signal. This is possible because extensive simulation work \cite{Kotake_2005zn} on core collapse SNe shows that the
onset of $\bar{\nu}_e$ luminosity is closely related to the time of the bounce.
%

\paragraph*{Master equation:}
Let us consider a gravitational detector and a neutrino detector with clocks
synchronized in universal time (U.T.). We  have:\footnote{Here and
in the following
times in uppercase are absolute times 
whereas times in lowercase are relative intervals of time.}
\begin{equation}
\tbounce=\tfirst - \tgwnue -\tmass  - \tfly - \tresp  \label{timerelation}
\end{equation}
%
where $\tbounce$ and $\tfirst$
are the absolute times, in U.T., of the bounce expected in the gravitational detector and of the first neutrino event
detected by the neutrino detector, respectively. 
The time $\tgwnue$ is the mean interval between the starting point of antineutrino luminosity and the bounce of the outer core on the
inner core. This is reliably known and ranges within $\tgwnue=(1.5-4.5)\mbox{ ms}$~\cite{Kachelriess2005}. The time $\tmass$ is the
delay, due to neutrino mass, between the arrival of GW
and neutrino signal;
however, this is limited by the cosmological bound $\sum_i m_{\nu_i} < 0.7 \mbox{ eV}$, that implies $ \tmass \sim 0.27
\left(\frac{m_{\nu}}{\mbox{\tiny 0.23 eV}}\right)^2 \left(\frac{\mbox{\tiny 10 MeV}}{E_{\nu}}\right)^2\left(\frac{D}{\mbox{\tiny 10 kpc}}\right)
\mbox{ms}$; thus, $\tmass$ appears negligible.
The time interval $\tfly$ is the time of fly between the
two detectors and depends on the SN position in the sky. 
Finally the non-negative parameter $\tresp$ is the difference of time between the first neutrino 
and the first event detected.
In summary, the main terms in Eq.~\ref{timerelation} are the fly time $\tfly$ and the response time $\tresp$; their quantitative evaluation is
discussed later.

By estimating the various terms in the right hand side of Eq.~\ref{timerelation}, we will determine the time of the bounce and the error in that
prediction. We note that $\delta \tbounce$ and $\delta \tfirst$ of the detector clocks are lower than $\mu s$; so their uncertainties can be
neglected for our purposes.

\begin{table}[t]
\begin{center}
{\footnotesize
\begin{tabular}{|c|c|c|c|c|c|c|}
  \hline
  & LIGO I & LIGO II & VIRGO & LVD & SK & IceCUBE   \\
  \hline
  $\Phi$  &  $30^\circ 30''$N& $46^\circ 27'$N& $43^\circ 41'$N & $42^\circ 28'$N & $36^\circ 14'$N & $90^\circ$S\\
  \hline
  $\lambda$ & $90^\circ 45'$W & $119^\circ 25'$W & $10^\circ 33'$E & $13^\circ 33'$E & $137^\circ 11'$E & $139^\circ 16'$W \\
  \hline
 $d^{\mbox{\tiny\sf SK}}$   & $32.1$ ms& $24.9$ ms & $28.8$ ms & $28.7$ ms & - &$19.0$ ms\\
 \hline
  $d^{\mbox{\tiny\sf LVD}}$    & $26.8$ ms & $27.5$ ms &  $0.9$ ms& -&  $28.7$ ms & $16.9$ ms\\
\hline
 $d^{\mbox{\tiny\sf IceCUBE}}$  & $20.8$ ms & $15.6$ ms &  $16.5$ ms& $16.9$ ms & $19.0$ ms&  - \\
    \hline
\end{tabular}}
\caption{\em \label{tab1} \footnotesize Coordinates  of 3 interferometers and 3 SN neutrino detectors and
temporal distances with Super-Kamiokande (resp., LVD and IceCUBE),
denoted by $d^{\mbox{\tiny\sf SK}}$ (resp., $d^{\mbox{\tiny\sf LVD}}$ and
$d^{\mbox{\tiny\sf IceCUBE}}$). \label{tab:tempidivolo}}
\end{center}
\end{table}
%

\paragraph*{Measuring $\tfly$:}

The time of fly between a neutrino and a GW detector
separated by the distance $\vec{d}$ 
is $\tfly=\vec{d} \; \hat{n}$,  
where $\hat{n}$ is the direction pointing to SN. The error is negligible for an astronomically identified SN. The same when we consider the
distance between LVD and VIRGO,  $d<1$~ms; in a sense, this is the ideal configuration. But the distances between Super-Kamiokande (SK) or
IceCUBE and the GW detectors LIGO or VIRGO are such (see Tab.~\ref{tab:tempidivolo}) that could imply an error as large as $\sim 60$~ms. Thus we
consider $\hat{n}$ as a random variable with most probable value $\hat{n}_*$ and find for $\delta\tfly^2=\langle \tfly^2\rangle-\langle
\tfly\rangle^2$:
\begin{equation}
\delta \tfly^2 = \left(\! d^{2} \!-\! (\vec{d}\; \hat{n}_*)^2 \! \right)\! \frac{\overline{\sin^2\!\theta}}{2} \! +\!  (\vec{d}\; \hat{n}_*)^2\!
 \left(\! \overline{\cos^2\!\theta} -(\overline{\cos\theta})^2 \!\right)
\label{eqer}
\end{equation}
where $\theta=\arccos(\hat{n}\; \hat{n}_*)$ is the angle of $\hat{n}$ with the SN direction. The first term typically dominates giving an error
$\delta \tfly\sim \delta \theta\times d$. Thus, to reach $\delta\tfly\le 5$ ms, we need to determine the angle with a precision of~$20^\circ$.

Tomas {\em et al.}~\cite{raff} remarked that
it is possible to do this with the elastic scattering  (ES) events of SK.
{\em E.g.}, consider a SN at 20 kpc. The search of the expected 35 forward ES events \cite{raff,prd} is simplified by minimizing the number of
inverse beta decay (IBD) events. These could be diminished by 20\% tagging the neutron \cite{nim2} and again by 20\%, requiring a visible energy
lower than 30 MeV \cite{strum,astr}. In fact, due to the neutrino in the final state, the ES events have a low average energy of $\sim 15$
MeV~\cite{astr} that means an angular resolution $\delta \theta=21^\circ$ \cite{nim}.
By simulating and then fitting the 
events we estimate the error in
the reconstructed direction. The average error on the angle is $5^\circ \pm 4^\circ$;
only 60 out of 10,000 simulations had a reconstructed angle larger than $20^\circ$,
occasionally due to a downward fluctuation of ES events. 
Thus, even in absence of an astronomical identification, it should be possible to determine the direction of the SN precisely enough to reduce
the error $\delta\tfly$ to the desired level. For a closer SN, larger number of ES events and/or better neutron identification, the measurement
will be safer.\footnote{To facilitate the search for the SN direction further, one could restrict the search to the galactic plane.}

\paragraph*{Measuring $\tresp$:}

\begin{figure}[t]
$$
\includegraphics[width=0.37\textwidth
]{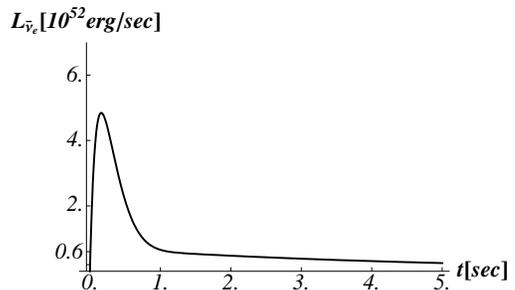}
$$
\caption{\em
The $\bar\nu_e$ 
luminosity in our model,
for the choice of parameters used to generate the events. The initial phase of
increased luminosity,  called ``accretion phase'' and connected
to the explosion 
is clearly visible.
\label{fig0}}
\end{figure}

The value of $\tresp$ and its uncertainty has to
be extracted from the data. If the astrophysical
mechanisms of neutrino emission were known
precisely the inference on the response time would be easy. Unfortunately this is not the case at present and we have to take into account the
astrophysical uncertainties. Thus we proceed as follows: First, we suppose that the expected flux of $\bar\nu_e$ from a standard core collapse
SN explosion can be described by a parameterized model; then, we fit at the same time the astrophysical parameters \emph{and} the response time
from the data.

We adopt and develop a model already
used for SN1987A data analysis \cite{noi}.
This model describes the $\bar{\nu}_e$ luminosity
from the instant when the shock wave,
originated from the bounce of the outer iron core on the
inner core of the star,
reaches the neutrino sphere and begins the neutrino emission,
until the end of the detectable neutrino signal.
%
The expression of the flux, whose luminosity is
depicted in Fig.~\ref{fig0}, 
is:
\begin{equation}
\Phi_{\bar\nu_e}(t) = f_r(t)\Phi_{a}(t) + (1-j_k(t)) \Phi_{c} (t-\tau_a).
\end{equation}
Here $t$ is the relative emission time, while
$\Phi_{a}$, $ \Phi_{c}$ and $j_k(t)$ are the accretion flux, the
cooling flux and the function
that links the two emission phases,
respectively.\footnote{The model is based on Section 3 of \cite{noi}.
Various source codes that implement this model can be downloaded
at the address
{\tt http://theory.lngs.infn.it/astroparticle/sn.html}.}
The expected rise \cite{Kachelriess2005} is described introducing:
\begin{equation}
f_r(t)= 1- 
e^{-t/\tau_r}
\end{equation}
that improves the existing parameterizations \cite{ll,noi}. The time scale $\tau_r \sim 50-300$ ms depends strongly \cite{Kotake_2005zn,janka
new} on the velocity of the shock wave; $\tau_r$ is the new, crucial model parameter. The accretion flux $\Phi_{a}$ is generated by the
interactions between the neutrons and the positrons above the shock and is described by 3 parameters: the initial accreting mass ($M_a$), the
time scale of the accretion phase ($\tau_a$), and the initial temperature of the $e^+$ ($T_a$). The cooling flux $\Phi_{c}$ coming from the
thermal emission of the new born proto-neutron star is proportional to the radius of the neutrino sphere ($R_c$), shows a time scale ($\tau_c$),
and an initial temperature of the emitted antineutrinos ($T_c$).
In summary, our parametrization of the $\bar\nu_e$ emission model includes 7 astrophysical parameters. 


\begin{figure}[t]
$$
\includegraphics[width=0.37\textwidth,height=0.2\textwidth]{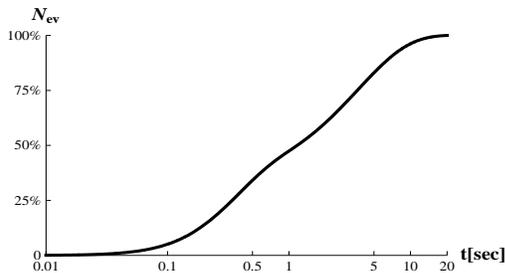}
$$
\caption{\em Curve of events accumulation in semilogarithmic scale. The slow (approximatively quadratic) initial accumulation is well visible.
The bump at $\sim 0.5$ s is due to the accretion phase. \label{fig01}}
\end{figure}

In order to construct a Monte Carlo simulation of a future SN event, we select the best-fit values of the parameters found from SN1987A data analysis \cite{noi}, namely
\begin{equation}
\begin{array}{cc}
R_c=16^{+9}_{-5} \mbox{ km}, & M_a=0.22^{+0.68}_{-0.15}\ M_{\odot},\\
T_c=4.6^{+0.7}_{-0.6}\mbox{ MeV}  & T_a=2.4^{+0.6}_{-0.4} \mbox{ MeV},\\
\tau_c=4.7^{+1.7}_{-1.2}\mbox{ s}, & \tau_a=0.55^{+0.58}_{-0.17}\mbox{ s},
\end{array}
\label{values}
\end{equation}
that are at odds with the theoretical expectations.
For the rise-time scale we choose the intermediate value $\tau_r=100\mbox{ ms}$ \cite{janka new}.
%
The expected IBD events rate is:
\begin{equation}
R(t,E_{\nu},D)= N_p\sigma_{\bar\nu_e p}(E_{\nu}) \Phi_{\bar\nu_e}(t, E_{\nu},D)\epsilon (E_{e^+}),
\end{equation}
where $D$ is the SN distance, $N_p$ is the number of target protons within the detector, $\sigma_{\bar\nu_e p}$ is the process cross section and
$\epsilon$ is the detector efficiency function. We show in Fig.~\ref{fig01} the cumulative curve for an energy threshold $E_{\mbox{\tiny
thr.}}=6.5\mbox{
  MeV}$ and constant detection efficiency.
We note that in the first 100 ms we expect to
accumulate 5\% of the total data set, this puts a limit on the
detector mass and/or on the SN distance needed to fit successfully
the parameter $\tau_r$ (as a thumb rule, we need
at least 20-30 events on average during the rise of the signal).

The total number of detected SN events is the integral of the rate function in the energy and in the detection time.
For a detection time window of $30$ seconds the number of expected events in a detector with the same mass of SK ({\em i.e.}, $22.5$ kton of
water) and efficiency $\epsilon=0.98$, is
\begin{equation}
N(D)=4233 \left(\frac{10 \mbox{ kpc}}{D}\right)^2 \textmd{   for  } E_{th}\geq 6.5\mbox{ MeV},
\label{numero}
\end{equation}
Thus, a SN neutrino burst from a galactic SN
will be unmistakably identified.

\begin{table}[t]
{\small
\begin{center}
\begin{tabular}{|c|ccccccc|}
\hline
$N_{\mbox{\tiny SN}}$ & $R_c$ & $T_c$  & $\tau_c$  & $M_a$  & $T_a$ & $\tau_a$  & $\tau_r$ \\[-1ex]
 & [km] & [MeV]  & [sec] & [$M_\odot$] & [MeV]  & [sec] & [ms] \\
\hline
$977$  & $14$ & $4.7$  & $4.6$ & $0.16$ & $2.4$  & $0.63$ & $51$\\
$1022$ & $15$ & $4.6$  & $4.8$ & $0.24$ & $2.3$  & $0.56$ & $86$\\
$1110$ & $14$ & $4.8$  & $4.7$ & $0.18$ & $2.4$  & $0.61$ & $99$\\
$1075$ & $15$ & $4.7$  & $4.6$ & $0.17$ & $2.5$  & $0.61$ & $79$\\
$1101$ & $16$ & $4.6$  & $4.7$ & $0.19$ & $2.4$  & $0.56$ & $104$\\
$1133$ & $15$ & $4.7$  & $4.8$ & $0.21$ & $2.4$  & $0.59$ & $69$\\
$1101$ & $16$ & $4.6$  & $4.8$ & $0.35$ & $2.3$  & $0.48$ & $166$\\
$1048$ & $16$ & $4.6$  & $4.6$ & $0.17$ & $2.5$  & $0.57$ & $100$\\
$1069$ & $16$ & $4.6$  & $4.7$ & $0.18$ & $2.5$  & $0.55$ & $126$\\
$1086$ & $17$ & $4.5$  & $4.8$ & $0.21$ & $2.5$  & $0.55$ & $172$\\
\hline
\end{tabular}
\end{center}}
\caption{\em Results of the analysis of ten simulated data sets for a SN event at $20$ kpc.
In the $1^{st}$ column there is the number of SN events extracted. In the subsequent six columns are the best-fit values for the astrophysical parameters. 
\label{tab:bestfit}}
\end{table}

Now we discuss the details of the analysis
procedure. We
extract a set of data from the rate function $R(t,E_{\nu},20)$,
expected for a SN event at $20$ kpc, that is a conservative or even pessimistic assumption.
Each event is characterized by
the relative detection time $t_i$ (namely the interval
time elapsed from the first detected event) and by the
positron energy $E_i$; the error on this energy
is obtained from the smearing function
$\delta E_i/E_i= 0.023 + 0.41\sqrt{\mbox{MeV}/E_i}$ \cite{nim}.
Finally we analyze the data set using a maximum likelihood
procedure to find the best-fit values of the 7 free astrophysical parameters of the emission model described above,
together with the $\tresp$ parameter,  the quantity that we want to estimate.
For each simulated data set we obtain from the fit a
value of this last parameter, that we call $\tresp^{\fit}$.
We will compare this fit value with the true value,
that we call $\tresp^{\true}$, and in this way, we will be able to
validate the procedure of analysis.


We show in Tab.~\ref{tab:bestfit} the best-fit values of the astrophysical parameters for ten simulated data sets, each one comprising
$N_{\mbox{\tiny SN}}$ data for a SN event at $20$ kpc and we compare them with the true values in Eq.~\ref{values}, used for the event
generator.
The comparison of the best-fit values for the astrophysical parameters can be used to test the
validity of the statistical procedure; it is remarkable that all these best-fit values are well
within the $1\sigma $ statistical errors found in \cite{noi} and reported in Eq.~\ref{values}.

\begin{table}[t]
\begin{center}
{\footnotesize
\begin{tabular}{|c|c|c|c|c|}
\hline
$\tresp^{\true}$& $\tresp^{\fit}$ & $|\tresp^{\true}-\tresp^{\fit}|$ & $2 \delta \tresp^{\fit}$  &$C$ \\[-1ex]
 {} [ms] &  [ms] & [ms] &  [ms]   & \\
\hline
$13$ & $6^{+6}_{-4}[1\sigma]^{+13}_{-6}[2\sigma]$    & $7$  & $9$  & 0.78 \\
$11$ & $7^{+14}_{-7}[1\sigma]^{+19}_{-13}[2\sigma]$   & $4$  & $22$  & 0.16\\
$9$ & $9^{+5}_{-4}[1\sigma]^{+13}_{-7}[2\sigma]$     & $0.3$   & $9$  & 0.03\\
$13$ & $5^{+4}_{-3}[1\sigma]^{+10}_{-5}[2\sigma]$ & $7$  & $7$  & 1.00 \\
$5$ & $7^{+5}_{-4}[1\sigma]^{+13}_{-6}[2\sigma]$ & $3$  & $9$  & 0.29 \\
$6$ & $5^{+4}_{-2}[1\sigma]^{+10}_{-5}[2\sigma]$ & $0.8$  & $6$  & 0.13 \\
$13$ & $5^{+5}_{-5}[1\sigma]^{+11}_{-9}[2\sigma]$ & $7$  & $10$  & 0.70 \\
$23$ & $11^{+7}_{-4}[1\sigma]^{+14}_{-8}[2\sigma]$ & $12$& $11$  & 1.10 \\
$3$ & $6^{+6}_{-3}[1\sigma]^{+13}_{-6}[2\sigma]$ & $2$  & $9$  & 0.29 \\
$2$ & $11^{+7}_{-4}[1\sigma]^{+16}_{-8}[2\sigma]$    & $9$  & $11$  & 0.85 \\
\hline
\end{tabular}}
\caption{\em Results of the ten simulations. The $1^{st}$ column are the true values of the response times,
the $2^{nd}$ column the estimated ones. In the $3^{rd}$ column we report the true error
and the $4^{th}$ column the $1\sigma$ estimated ones. In the last column we show
the values of the compatibility error factor. 
}\label{tab:tbounce}
\end{center}
\end{table}

The results for $\tresp^\fit$ are given in  Tab.~\ref{tab:tbounce}.
In the first column there are the true values 
of the response time $\tresp^\true$, 
namely the interval of time between the first neutrino
detected and the first neutrino arrived in the detector. 
In the second column are the corresponding 
best-fit values as determined from the
maximization of the likelihood of the simulated data set and the statistical errors found by Gaussian procedure. The third column shows the
difference between the true value and the estimated one, namely the {\em true} error of our procedure. The fourth column gives the $1\sigma$
range of error, $2 \delta \tresp^\fit$, as evaluated from 
the second column. 
This is compared with the true error in the fifth column, by means
of the compatibility error factor:
\begin{equation}
C=\frac{|\tresp^\true-\tresp^\fit|}{2 \delta \tresp^\fit}. \label{compatibility}
\end{equation}
When this is lower than 1 the compatibility is good and the $1\sigma$ statistical error can be used to find the true value of the response time.
The results show that this is the case. Thus, we can estimate the 
true time of the bounce with an average uncertainty time window of $\langle{2
\delta \tresp^\fit} \rangle= 10.5\mbox{ ms}$.

\paragraph*{Summary:} 
Summing in quadrature the errors of the terms 
in Eq.~\ref{timerelation}, 
the time of the bounce can be located into a temporal window of
about $15$~ms for a SN at $20$kpc.

\paragraph*{\bf Discussion:}
A galactic SN will permit us to obtain very detailed information on the time structure of the neutrino burst, thanks to large detectors as SK
(capable to indentify the direction of the SN even in absence of an astronomical observation), thanks to a lucky configuration between LVD-Virgo
(practically in the same location) and possibly  thanks to new detectors such as IceCUBE.\\ \indent We have shown that even in rather
conservative assumptions,  namely for a very distant galactic SN, it will be possible to use the neutrino data to predict the time of the burst
of gravity waves with a precision comparable to its expected duration. More in detail, the use of Eq.~\ref{timerelation} allows the
determination of the time of the bounce with a precision of few tens of milliseconds even for a galactic SN exploding at a distance of 20 kpc
from us. While the proposed method mostly relies on the analysis of the conventional inverse beta decay events, we have argued that the elastic
scattering (ES)
events detected by SK could add precious information.\\
\indent However, this type of analysis can be useful {\em even if} a sample of ES events cannot be precisely identified. Indeed, the large
number of events detected by SK and IceCUBE allows us to deduce the astrophysical parameters that describe the observable neutrino
signal, including the most crucial one, namely the rise-time $\tau_r$.
This information, inserted as a ``prior'' in the analysis of LVD data, greatly
enhances the capability of our procedure to deduce
with good precision $\tresp$
from the relatively smaller LVD data set. The response time,
determined in this way, can be used as a reliable
trigger for the search of GW in VIRGO.

{\tiny
We thank V.~Fafone, E.~Katsavounidis, K.~Scholberg, M.~Selvi, F.L.~Villante\\[-1ex]
for useful
discussions. Preliminary results presented by G.P.~at GWADW 2008,
\\[-1ex]
Isola d'Elba and at the
ILIAS
meeting, Cascina, by
F.V.~at the 2008 SIF meeting,\\[-1ex]
and in the PhD thesis
of G.P., L'Aquila University, submitted on Dec.~2008.\\[-1ex]
Partly supported by High
Energy Astrophysics Studies contract no.~ASI-INAF\\[-1ex]
I/088/06/0,
MIUR grant for
the Projects of National Interest
PRIN 2006 ``As-\\[-1ex]
troparticle Physics'', FP6
Eur.\ Network ``UniverseNet'' MRTN-CT-2006-035863.}

\end{document}